\documentclass[12pt,a4paper]{extarticle}
\usepackage[utf8]{inputenc}
\usepackage[T1]{fontenc}
\usepackage[a4paper, left = 2.5cm,right = 2.5cm,top = 2.5cm,bottom = 3cm]{geometry}
\usepackage{amsmath,amssymb}
\usepackage{graphicx}
\usepackage{lmodern}
\usepackage{cite}
\usepackage[linktocpage=true, colorlinks=false, linkcolor = blue, citecolor = red]{hyperref}
\begin{document}
	\title{Global embedding of BTZ spacetime\\ using generalized method \\of symmetric embeddings construction}
	
	\author{A. A. Sheykin\thanks{\texttt{anton.shejkin@gmail.com}}, \ M. V. Markov\thanks{\texttt{vkrms321@gmail.com}}, \ S. A. Paston\thanks{\texttt{pastonsergey@gmail.com}}\\
		\textit{Saint Petersburg State University, St. Petersburg, Russia}}
	\date{}
	\maketitle
	\abstract{It is often easier to study pseudo-Riemannian manifolds by presenting them as surfaces in some ambient space. 
		We propose an algorithm for construction of explicit isometric embeddings of pseudo-Riemannian manifolds with symmetries into an ambient space of higher dimension. While most of the existing methods are based on Gauss-Codazzi-Mainardi-Peterson equations, we do not use them and instead concentrate on a system of equations which connects the metric on the manifold and the embedding function of the surface. Our algorithm is based on the group theoretical method of separation of variables that we developed earlier (arXiv:1202.1204). The algorithm makes this method more convenient and simple to use. It allowed us to simplify the construction of many known embeddings as well as obtain some new ones. In particular, we obtain explicit global embeddings of spinning BTZ black hole in 7-dimensional flat space.}
	\section{Introduction}
	The study of curved surfaces has always been extremely helpful in the understanding of non-Euclidean geometries. After the seminal works of Gauss \cite{gauss1902} and Riemann \cite{Darrigol} it became clear that these strange geometries can be treated simply as geometries of curved surfaces in a flat ambient space, with whom the mathematicians have been familiar since ancient times. In particular, Gauss obtained a formula which provided a decomposition of covariant derivatives of a vector tangent to a surface into tangent and normal parts. This formula now bears his name. 
	
	Several other prominent mathematicians also made their contributions to the theory of curved surfaces. In 1861 Weingarten obtained an analogous equation for the derivative of the normal vector, so the full system of equations is called Gauss-Weingarten equations \cite{taimanov2008lectures}. Gauss proved his famous \textit{Theorema Egregium} from which it follows that the Gaussian curvature of a surface is invariant w.r.t. its isometries, i.e. it is completely determined by its first fundamental form (also known as metric). Bonnet proved that first and second fundamental form of the surface (with the latter, at least for 2-surface in 3-space, can intuitively be imagined as the matrix constructed from coefficients of Taylor expansion of surface equation), define a surface in an unique way, up to possible motions in an ambient space. 
	
	However, not each vector field can serve as a vector field of normals to some surface, so certain integrability conditions are needed. The system which provides these conditions is called Gauss-Codazzi-Mainardi-Peterson equations. It was obtained for a first time by Peterson in 1853 (who also proved Bonnet theorem earlier than Bonnet himself), but this fact had been recognized relatively lately \cite{Phillips}. \footnote{It should be noted that GCMP equations are often supplemented by the so-called Ricci equations for the sake of convenience, although these Ricci equations are not completely independent from GCMP ones, see \cite{Goenner1977} and \cite{slemrod} for details.} This complicated system of quasi-linear PDEs is usually considered as the main object of study in the geometry of curved surfaces. If one wants to determine whether a curved space with given metric can be locally isometrically embedded into an ambient space, one could solve GCMP equations with respect to second fundamental form (or at least prove that it has a solution). A survey of results and methods related to the GCMP equations can be found, e.g., in \cite{schmutzer,slemrod,hanhong}.
	
	This method of searching for an embedding, however, has its drawbacks. Firstly, it does not give an explicit parametric form of the surface. Although it is possible to reconstruct it from the second fundamental form, this procedure is not a completely trivial one. Secondly, it works well only in cases when the codimension of a surface (i.e. the difference between dimensions of the ambient space and the surface) is small. In fact, the second fundamental form was initially called the second \textit{quadratic} form, since in the codimension 1 it indeed resembles a quadratic form such as the metric. Since the early works were mainly devoted to the embeddings of 2-surfaces in 3-space, initially this circumstance was not a big issue, but the complexity of calculations raises dramatically with the increasing codimension, and very little is known about solutions of such system if the codimension exceeds 2. Lastly, it does not straightforwardly allow to use the presence of some symmetry in the metric which is to be embedded (although some ideas about its usage in the analysis of GCMP equations are present in the literature, see, e.g., \cite{Goenner1975}, \cite{Chantry_2021} and references therein).  
	
	Since in several physical applications it is often necessary to know an exact parametric form of the embedded surface, the codimension of such surface in some cases can be quite high, and the physically interesting metrics often possess a rich symmetry group, it seems plausible to explore the alternative methods of construction of isometric embeddings. One of such methods was invented \cite{statja27} in order to construct and classify all possible embeddings of Schwarzchild black hole metric. Later it was successfully applied to the embeddings of non-spinning black holes of various kinds in 3+1 \cite{statja30,statja32,statja40,statja57} and 2+1\cite{statja56} general relativity, and to some cosmological solutions \cite{statja29,2004.05882}. In the present paper we propose a modification of this method which allows to search for embeddings of more complicated spacetimes (e.g., spacetimes with spinning sources). To illustrate the possibilities of this method, we construct several explicit embeddings of metrics with symmetries.
	
	This paper is organized as follows. In Section \ref{2} we outline basic properties of explicit embeddings into a flat space and briefly review the original method of symmetric embeddings construction. In Section \ref{3} we describe a generalization of this method based on a reduction of the symmetry group of the metric, and give some simple examples of its application. In Section \ref{4} we construct explicit embeddings of BTZ black hole and G\"{o}del universe.
	This 7-dimensional embedding of BTZ black hole turns out to be global, so it is considerably better that 10-dimensional local embedding described in the literature \cite{Willison_2011}.
	Finally, in the last section we make some comments about future possible modifications and applications of this method.

	\section{Embedding of symmetric manifolds}\label{2}
	\subsection{Some basic formulas and conventions}
	The main theorem which determine the characteristics of ambient space in which the surface with given metric can be isometrically embedded, is the following:
	
	\textit{An arbitrary $n$-dimensional (pseudo)-Riemannian mainfold with analytic metric can be locally isometrically embedded into an $N$-dimensional flat ambient spacetime with $N=\frac{n(n+1)}{2}$.} \footnote{There are some trivial remarks that we want to make for the sake of completeness: first, the ambient spacetime must contain at least the same number of spacelike and timelike directions (i.e. the metric of ambient space must not have less positive and negative eigenvalues than the metric of the manifold itself); and second, embeddings into spacetime with more than $n(n+1)/2$ dimensions are of course also allowed.}
	
	For spaces with positively defined metrics it was posed as a conjecture by Schl\"afli in 1873. The outline of the proof was given by Janet in 1926 \cite{gane} (a complete proof based on Janet's ideas was carried out by Burstin in 1931 \cite{burstin}, who also extended it on embeddings \textit{into} a Riemannian space with analytic metric); and in the subsequent year, Cartan gave an independent proof \cite{kart}. In 1961 Friedman \cite{fridman61} proved that the same dimension is required in a pseudo-Riemannian case. 
	
	The number of dimensions required for a global embedding of an arbitrary $n$-dimensional surface is drastically higher than $n(n+1)/2$ (for a review, see \cite{Gromov_1970}). On the contrary, in many particular cases a specific $n$-dimensional pseudo-Riemannian manifold can be globally isometrically embedded into a spacetime which dimension is much lower than $n(n+1)/2$, e.g. $n$-spaces of constant curvature can always be globally embedded into $(n+1)$-space. The minimal possible codimension of an isometrically embedded surface with given metric is called an \textit{embedding class}: $p=N_{\min}-n$. It can serve as an invariant characteristic of a spacetime and therefore is often used for the classification purposes \cite{schmutzer}.
	
	The exact form of the surface can be defined in both implicit and explicit ways: implicit definition puts a set of constraints on a set of ambient space coordinates $y^a$, $a=1\ldots N$:
	\begin{align}
		F^A (y^a)=0,  
	\end{align}
	where $A=1\ldots (N-n)$.
	It is useful in cases when one have no opportunity or desire to introduce a coordinates on the surface (one of coordinate-free approaches to the theory of embedded surfaces can be found, e.g., in \cite{statja25,statja61}). However, if a coordinate map $x^{\mu}$ is already chosen, e.g., by metric structure, the explicit way of surface definition is much more convenient. It consists of defining an \textit{embedding function}:
	\begin{align}
		y^a=y^a(x^\mu),
	\end{align}
	$\mu=0\ldots (n-1)$.
	Using the fact that the ambient space is also equipped with a metric $\eta_{ab}$, we can write
	\begin{align}
		g_{\mu\nu} (x) dx^\mu dx^\nu = ds^2 = \eta_{ab} dy^a (x) dy^b (x) = \eta_{ab} \frac{\partial y^a}{\partial x^\mu} \frac{\partial y^b}{\partial x^\nu} dx^\mu dx^\nu  
	\end{align}
	to obtain a formula for \textit{induced metric}:
	\begin{align}\label{metric}
		\frac{\partial y^a}{\partial x^\mu} \frac{\partial y^b}{\partial x^\nu} = g_{\mu\nu} (x), 
	\end{align}
	so the first fundamental form of the surface can be expressed in terms of the embedding function. The abovementioned second fundamental form can also be obtained as a second covariant derivative of $y^a$:
	\begin{align}
		b^a_{\mu\nu} = D_\mu D_\nu y^a,
	\end{align}
	so it can be concluded that the maximal amount of information about both intrinsic and extrinsic geometry of an embedded surface is encoded in the embedding function.
	
	This function, however, is quite hard to find. To obtain an explicit form of $y^a$, one has to solve a system \eqref{metric}, which is a system of nonlinear PDEs. Such systems are practically unsolvable in the general case, and the methods of solving them are scarce (a collection of results related to the solutions of the system \eqref{metric} can be found in \cite{jensen}). The solution is much simpler to find when the metric at the r.h.s. of \eqref{metric} possesses some symmetries. The method of simplification based on symmetries of the metric is a topic of the next subsection. 
	\subsection{Original method}
	The symmetry-based method of construction of explicit embeddings was proposed in \cite{statja27}. We will briefly outline it here for the sake of convenience and refer the reader to the original paper for details. 
	
	First of all, let us note that the system \eqref{metric} becomes much simpler to solve if it allows a complete separation of variables. Such separation can be achieved in case when the dependence of embedding function on all coordinates (possibly except one) is determined by symmetry restrictions. Let us describe how to put such restrictions on it.
	
	In order to connect the notions of symmetry in terms of intrinsic and extrinsic geometry let us define a \textit{symmetric surface}:
	
	\textit{A surface $\mathcal M$ is called symmetric w.r.t. group $G$ if $\mathcal M$ transforms into itself under the action of some subgroup $\hat{G}$ of the group of motions $\mathcal P$ of an ambient space isomorphic to $G$: $\hat{G}\sim G$.}
	
	Therefore a homomorphism $V$ exists which maps the symmetry group $G$ of the metric into the group of motion  $\mathcal P$ of the ambient space: $V: G \to \mathcal P$. Then if one takes a \textit{generatrix} --- some vector in ambient space, and then transforms it with matrices of some representation of $V(g)$, then the resulting surface would have a symmetry group $G$, and its dependance of those coordinates that are affected by $G$ would be fixed.
	
	Therefore this definition suggests a procedure of constructing of surfaces whose metric has a symmetry group $G$: one could count all suitable representations of $G$, starting with the lowest possible dimension, and pick out those that looks like some subgroups of $\mathcal P$. Then one should choose a generatrix $y_0$ for each such representations $V(g)$, and the desired surfaces can be obtained as a result of acting of representations $V$ on $y_0$. This is the general scheme which works for any $G$ and $\mathcal P$ (assuming, of course, that $\mathcal P$ is sufficiently large to contain a subgroup isometric to $G$, otherwise our definition of symmetric surface obviously would not work), though the details can vary depending on dimensionality and globality of desired embedding. In the next subsection we will discuss a specific example of this procedure: the construction of surfaces with an Abelian symmetry group embedded into Minkowski ambient spacetime.  
	
	\subsection{Surfaces with Abelian symmetry in Minkowski spacetime}\label{zz}
	To begin with, let us deal with the suitable form of realizations of $\mathcal P$, which for $(m+n)$-dimensional Minkowski spacetime is the Poincare group:
	\begin{align}
		\mathcal{P}_{m,n} =  \text{SO}(m,n)\ltimes \mathbb{R}^{m,n}.
	\end{align}
	It is known \cite{Hall2015} that $\mathcal{P}_{m,n}$ can be represented by the following matrices: 
	\begin{align} \label{Pmn}
		\mathcal{P}_{m,n} \sim 	\begin{pmatrix}
			\Lambda & a\\ 0 & 1
		\end{pmatrix}, \ \Lambda \in \text{SO}(m,n), \ a \in \mathbb{R}^{m,n}
	\end{align}
	Therefore we are interested in such representation of metric symmetry group $G$ that looks like \eqref{Pmn}.
	
	Now let us deal with representations of an Abelian group. It is known that an arbitrary matrix representation of the one-parametric Abelian group with parameter $t$ can be written as $V(t) = \exp(\omega t W)$, where $\omega$ is a constant with dimension of $t^{-1}$ (it can be omitted if $t$ is dimensionless), and $W\in M_{n\times n} (\mathbb{C}), n\in \mathbb{N}$. Since any matric can be brought to its Jordan form, we can conclude that $W$ consists of Jordan blocks of the form 
	\begin{align}
		W=\begin{pmatrix}
			\lambda    & 1      & 0      & \dots \\
			0      & \lambda    & 1      & \dots \\
			\vdots  & \vdots  & \ddots  & \ddots \\
			0      & 0      & 0      & \lambda   \\
		\end{pmatrix},
	\end{align}
	where $\lambda\in \mathbb{C}$. Decomposing $\lambda=(\beta+i\alpha)/\omega$, we can write
	\begin{align}\label{V_}&V(t)=S_s(\omega t)e^{i\alpha t}e^{\beta t},\end{align}
	{where}
	\begin{align}\begin{split}
			\qquad	S_0=1&,\
			S_1=\left(\begin{matrix}
				1 & \omega t\\
				0 & 1
			\end{matrix}\right),\
			S_2=\left(\begin{matrix}
				1 & \omega t& {(\omega t)^2}/{2!}\\
				0 & 1 &  \omega t\\
				0 & 0 & 1
			\end{matrix}
			\right), \\
			&S_3=\left(\begin{matrix}
				1 & \omega t& {(\omega t)^2}/{2!} & {(\omega t)^3}/{3!}\\
				0 & 1 &  \omega t & {(\omega t)^2}/{2!}\\
				0 & 0 & 1 & \omega t\\
				0 & 0 & 0 & 1
			\end{matrix}\right),\ldots 
		\end{split}
	\end{align}
	Let us consider a block diagonal matrix whose elements resembles various blocks in \eqref{V_} in order to obtain the representations of the form \eqref{Pmn}. 
	
	The sufficient condition for a matrix $M$ to belong to $\mathcal{P}_{m,n}$ is that it must preserve some symmetric bilinear form $\eta$: $M^T \eta M = \eta$. Firstly, one can take a realification of $e^{i\alpha t}$ to obtain
	\begin{align}\label{R}
		R(\alpha) = \begin{pmatrix}
			\cos \alpha t & \sin \alpha t\\
			-\sin \alpha t & \cos \alpha t,
		\end{pmatrix},
	\end{align}
	which is a rotation in a Euclidean plane conserving the metric
	\begin{align}
		\eta = \begin{pmatrix}1 & 0\\ 0 & 1\end{pmatrix}
	\end{align}
	so $R(\alpha)\in \mathcal{P}_{m,n}$. Then one can take a direct sum of two one-dimensional representations $e^{\beta t}$ and $e^{-\beta t}$ to obtain \begin{align}
		B'(\beta) = \begin{pmatrix}
			e^{\beta t} & 0\\
			0 & e^{-\beta t}\end{pmatrix}
	\end{align}
	which is a boost matrix in light cone coordinates preserving a metric
	\begin{align}
		\eta' = \begin{pmatrix}0 & 1\\ 1 & 0\end{pmatrix}.
	\end{align}
	By change of coordinates one can rewrite it in a more familiar form:
	\begin{align}\label{B}
		B(\beta) = \begin{pmatrix}
			\cosh \beta t & \sinh \beta t\\
			\sinh \beta t & \cosh \beta t 
		\end{pmatrix}, \quad \eta = \begin{pmatrix}
			1 & 0\\
			0 & -1 
		\end{pmatrix}.
	\end{align}
	Now let us deal with the $S$-matrices. The easiest one is probably $S_1(\omega t)$, which is simply a matrix of translation in a $t$ direction, whereas the most tricky one is $S_2$, which at first glance does not look familiar at all. However, a direct calculation shows that $S_2$ preserves the following metric:
	\begin{align}\eta=\begin{pmatrix}
			0 & 0& 1\\
			0 & -1 & 0\\
			1 & 0 & 0
	\end{pmatrix}\end{align} 
	and thus indeed belongs to $\mathcal{P}_{m,n}$. This transformation is known under several names, e.g., Vilenkin called it \textit{parabolic rotations} \cite{vilenkin}. Other $S$-matrices with even index $S_{2k}$ can also be considered as rotations of special kind, whereas $S$-matrices with odd index $S_{2k+1}$ can be considered as $S_{2k}$-rotations accompanied by a translation as all $S_{2k+1}$ clearly have the structure of \eqref{Pmn}. 
	
	Note that it is possible to combine more than one realization of $t$-shifts to obtain representations with higher dimension. Suppose we are interested in embedding of 2-dimensional $t$-symmetric spacetime into a 3-dimensional flat space:
	\begin{align}
		ds^2 = g_{00}(r) dt^2 - 2 g_{01}(r) dt dr - g_{11}(r) dr^2.
	\end{align}
	Then it is possible, for example, to combine translations with rotations, which gives a spiral-like surface in 3-dimensional space. The resulting representation then has the form $V(t) = R(\alpha)\oplus S_1(\omega t)$. To ensure that the surface is 2-dimensional, one can choose the generatrix as 
	\begin{align}
		\begin{pmatrix}
			y_0\\1
		\end{pmatrix} = 	\begin{pmatrix}
			f(r)\sin (g(r))\\f(r)\cos (g(r))\\h(r)\\1
		\end{pmatrix},
	\end{align}
	and the resulting $t$-symmetric surface takes the form:
	\begin{align}\label{spiral}
		\begin{pmatrix}
			y\\1
		\end{pmatrix}  = 
		\begin{pmatrix}
			\cos(\alpha t) & \sin(\alpha t) & 0 & 0\\
			-\sin(\alpha t) & \cos(\alpha t) & 0 & 0\\
			0& 0& 1 & \omega t\\
			0&0&0&1 
		\end{pmatrix} \begin{pmatrix}
			f(r)\sin (g(r))\\f(r)\cos (g(r))\\h(r)\\1
		\end{pmatrix} = 
		\begin{pmatrix}
			f(r)\sin (\alpha t + g(r))\\f(r)\cos (\alpha t +g(r))\\\omega t + h(r)\\1
		\end{pmatrix}.
	\end{align}
	Since the dependence on $t$ is fixed now, one can plug this embedding function into \eqref{metric} and solve the system of ODEs w.r.t. $f(r), g(r)$ and $h(r)$. This spiral-like surface \eqref{spiral} is one of the six types of $t$-symmetric 2-dimensional surfaces in 3-dimensional ambient spacetime, which are elliptic, parabolic and hyperbolic surfaces with or without accompanying translations. It is interesting to note that similar results were obtained in the classification of stationary trajectories in 4-dimensional Minkowski spacetime (i.e. 1-dimensional $t$-symmetric surfaces) \cite{letaw81}, although this analysis was not based on group theory.

	Note also that in case when the Abelian symmetry group of the metric is compact (e.g., $\text{SO}(2)$), all these types of representations except $R(\alpha)$ would give only local embeddings since they correspond to non-compact surfaces.
	
	This is an example of application of the original method which will be useful later.
	Other symmetry groups, including non-Abelian ones, can be considered in the framework of this method (see \cite{statja27} for a detailed study of $SO(3)$-symmetric surfaces and \cite{statja29} for $SO(4)$, $SO(1,3)$ and $SO(3)\ltimes \mathbb{R}^3$).
	\section{Symmetry reduction}\label{3}
	The method described above has proved itself useful in many cases. It has, however, its limitations. Namely, a complete classification of the representations of a given group with desired dimension can be quite tedious and gives a lot of possible surface types which then needs to be checked out manually. Furthermore, as the method prescribes a complete matching of internal and external symmetry, it does not allow to construct surfaces whose symmetry is lower than the symmetry of metric on them. Therefore one could search for possible generalization of this method. The idea of such generalization was for the first time proposed in \cite{2004.05882}: it was noticed there that for manifolds with large symmetry group it could be sufficient to use only a part of it to separate the variables in \eqref{metric} completely. This idea worked well, e.g., for spaces of constant curvature as well as for various cosmological models which also have high symmetry. In this section we will describe an algorithm for embedding construction based on the idea of symmetry reduction.
	\subsection{Diagonal metrics}
	The idea of the procedure is to use the maximal number of Abelian subgroups containing in the symmetry group of the metric. It is known that one can associate a Killing vector with each Abelian subgroup of the metric, and the presence of such Killing vector at the level of metric manifests itself in the fact that metric components are independent of the corresponding coordinate (note that coordinate system needs to be chosen in such way that a translation w.r.t. certain coordinates would correspond to a Killing vector). It is sufficient to show the procedure in case when only one such vector is present, as they are supposed to commute with each other and can be dealt with separately. Therefore the metric of such spacetime can be written in the following form:
	\begin{align}\label{genmet}
		ds^2= g_{ij}(x^1,...,x^n)dx^idx^j+g_{\phi\phi}(x^1,...,x^n)d\phi^2+2g_{\phi i}(x^1,...,x^n)d\phi dx^i,
	\end{align}
	where $\phi$ is a coordinate associated with Killing vector and indices $i,k,\ldots$ take all values except $\phi$.
	Let us also assume that $g_{\phi i}=0$ (the case $g_{\phi i}\neq 0$ will be considered in the next Section). Our aim is to determine the dependence of the embedding function on the coordinate $\phi$ in order to simplify the system \eqref{metric}. To do that, it is sufficient to construct an embedding of $g_{\phi\phi}$. This means that one could try to choose a form of some components of embedding function in such way that their contribution to induced metric would be equal to $g_{\phi\phi}$, non-diagonal metric components w.r.t. $\phi$
should not appear after that and none of components of the metric of the remaining submanifold should depend on $\phi$.
	It is easy to see that in general case it is not possible to achieve with only one component, so we need at least two of them (or more: in fact, all realizations that are listed in the Section \ref{zz} are suitable as long as we are not bounded by restriction of dimensionality or periodicity).
	
	As we know from the previous section, two-dimensional representations of the translational symmetry can either be of the form of rotations \eqref{R} or boosts \eqref{B}. They can be put together in the following ansatz:  
	\begin{align}   	\label{alg}\begin{split}
		y^1=\sqrt{\xi}&\dfrac{\sqrt{g_{\phi\phi}}}{a}\sin \sqrt{\xi} a\phi,\\
		y^2=\phantom{\sqrt{\xi}}&\dfrac{\sqrt{g_{\phi\phi}}}{a}\cos \sqrt{\xi} a\phi.
	\end{split}\end{align}
	where $a$ is a constant, $\xi=\pm1$ and the signature must be $(1,\xi)$.
	The sign of $\xi$ should be positive if $\phi$ is compact (so-called elliptic embedding) and negative otherwise (hyperbolic embedding). In the formula \eqref{alg} it is assumed that $g_{\phi\phi}\geq0$, and if it is not, the sign conventions and the signature of ambient spacetime needs to be adjusted correspondingly. If $g_{\phi\phi}$ changes its sign at one point, one should choose hyperbolic embedding and perform a junction of two embeddings with different order of hyperbolic functions at this point, see details in \cite{statja27}. After that the metric which has to be embedded takes the form
	\begin{align}\label{ds}
		d\tilde{s}^2 = \tilde{g}_{ij} dx^i dx^j = g_{ij}(x^1,...,x^n)dx^idx^j-\xi\dfrac{(d g_{\phi\phi})^2}{4 a^2 g_{\phi\phi}}
	\end{align}
	As we said earlier, if there are more than one $\phi$-like coordinate, the above steps should be repeated for all of them. If after accounting of all coordinates affected by symmetry the remaining spacetime to be embedded with the metric $\tilde{g}_{ij}$ turns out to be one-dimensional (i.e. there is only one coordinate $x$ left), it can be embedded in 1-dimensional spacetime
	\begin{align}\label{int}
		y=\int dx \sqrt{\pm\tilde{g}_{xx}}.  
	\end{align}
	If the obtained embedding lacks  globality, which usually happen when $\tilde{g}_{xx}$ can change sign or has poles, one could step back to the situation when there is one symmetric coordinate in addition to $x$ and try to embed 2-dimensional space $(\phi,x)$ into 3-dimensional space via \eqref{spiral} or other types mentioned earlier.

	It should be noted that the method described here also allows to construct isometrically bendable surfaces, i.e. surfaces that can be continuously deformed without the alteration of their metric. The constant $a$ in \eqref{alg}  plays the role of a bending parameter. Such deformation, however, often leads to the loss of the globality of an embedding due to the appearance of conical singularities.
	
	\subsection{Almost diagonal metrics} 
	In this section we consider the embedding of metrics which contain a non-diagonal term:
	\begin{multline}
		ds^2= g_{ij}(t,x^1,...,x^{n-1})dx^idx^j+\\+g_{\phi \phi}(t,x^1,...,x^{n-1})d\phi^2+2g_{t \phi}(t,x^1,...,x^{n-1})dtd\phi,
	\end{multline}
	where $g_{t \phi}\neq 0$, $\phi$ is a symmetric coordinate and $t$ is an arbitrary one, $x^i=\{t,x^1,...,x^{n-1}\}$. Here and hereafter we will consider only one term $g_{t\phi}$ for the sake of simplicity, since the procedure can be applied to each non-diagonal term separately due to the fact that by construction it can only reduce the overall number of non-diagonal terms and does not lead to the appearance of new ones.
	
	In the particular case, when $g_{t\phi} = f(t) g_{\phi\phi}$, we can deal with $g_{t \phi}$ in a way similar to $g_{\phi\phi}$, embedding it into 2-dimensional ambient space with signature $(1,\xi)$, and coordinates
	\begin{align}\begin{split}\label{ura}
			&y^1=\sqrt{\xi}{\sqrt{\frac{g_{t\phi}}{b h'(t)}}}\sin \sqrt{\xi} (b\varphi+h(t)),\\
			&y^2=\phantom{\sqrt{\xi}}{\sqrt{\frac{g_{t\phi}}{b h'(t)}}}\cos \sqrt{\xi} (b\varphi+h(t)),\end{split}
	\end{align}
	where $h(t)$ has the form
	\begin{align}\label{h}
		h(t)=\int dt \frac{b g_{t\phi}}{g_{\phi\phi}}  = 
		b \int dt f(t).
	\end{align}
	This embedding allows to get rid both of $g_{t\phi}$ and $g_{\phi\phi}$ using just two components of $y^a$.
	
	When the dependence of in $g_{\phi\phi}$ and $g_{t\phi}$ on other coordinates is more complicated, it is no longer possible to embed both of them using just one block, and $g_{\phi\phi}$ needs to be embedded separately, as prescribed by \eqref{alg}. To embed  $g_{t\phi}$, it is convenient to choose 
	\begin{align}
		h(t)=a t,
	\end{align}
	so \eqref{ura} transforms into
	\begin{align}\begin{split}\label{nond}
		&y^1=\sqrt{\xi}\sqrt{\dfrac{g_{t \phi}}{ab}}\sin(\sqrt{\xi}(b\varphi+at)),\\
		&y^2=\phantom{\sqrt{\xi}}\sqrt{\dfrac{g_{t \phi}}{ab}}\cos(\sqrt{\xi}(b\varphi+at)),\end{split}
	\end{align}
	where $a$ and $b$ are positive constants and signature is $(1,\xi)$ with $\xi=\pm1$. As in \eqref{alg}, here we assume that $g_{t \phi}\geq 0$, if it is not, signature conventions and signs of $a$ and $b$ must be changed. Direct calculation shows that the metric that remains to be embedded does not contain non-diagonal components w.r.t $\phi$ and thus can be embedded using the method described in the previous Section.
	
	\subsection{Example: a particle in 2+1 dimensions}
	Now let us illustrate this method by a simple example from 2+1-dimensional GR. It is a well-known fact that gravitational field in an empty 2+1-dimensional spacetime does not have local degrees of freedom, so an addition of point sources do not affect the metric of the spacetime. It leads, however, to some global effects which could easily be described as isometric bending of corresponding surface. Let us apply the generalized method described above to the metric of a 
	2+1 dimensional spacetime of constant positive curvature\cite{deserjackiw}:
	\begin{align}\label{la}
		ds^2=\cos^2 \theta \, dt^2-\dfrac{1}{\Lambda}(d\theta^2+\sin^2\theta \, d\phi^2),
	\end{align}
	where $\Lambda>0$ is a cosmological constant and the parametrization $(t,\theta,\phi)$ is analogous to Hopf parametrization of a 3-sphere. The symmetry group of this metric is $SO(1,3)$, but we will not be interested in the construction of $SO(1,3)$-symmetric surface. Instead of this, let us apply our generalized method. There are two Abelian subgroups of $SO(1,3)$: $SO(2)$ associated with $\phi$ and $SO(1,1)$ associated with $t$. Thus the components $g_{\phi \phi}$ and $g_{tt}$ can be embedded using the following ansatz:
	\begin{align}\begin{split}\label{L>0 emb}
		y^1=\dfrac{1}{a\sqrt{\Lambda}}\cos\theta\sinh a \sqrt{\Lambda}  t,\phantom{\Bigg( } \quad & y^3=\dfrac{1}{b \sqrt{\Lambda}  }\sin\theta\sin b\varphi,\phantom{\Bigg( }\\
		y^2=\dfrac{1}{a\sqrt{\Lambda}}\cos\theta\cosh  a \sqrt{\Lambda} t, \quad & y^4=\dfrac{1}{ b\sqrt{\Lambda} }\sin\theta\cos b\varphi. \end{split}
	\end{align}
	where $a$ and $b$ are constants and signature is $(+,-,-,-)$
	The remaining metric component
	\begin{align}\label{pyat}
		\tilde{ds^2}=-\dfrac{1}{\Lambda}\left(1-\dfrac{\sin^2(\theta)}{a^2}-\dfrac{\cos^2(\theta)}{b^2}\right)d\theta^2,
	\end{align}
	can be embedded using just one additional spacelike component:
	\begin{align}\label{p}
		y^5=\dfrac{1}{\sqrt{\Lambda}}\int \sqrt{1-\dfrac{\sin^2(\theta)}{a^2}-\dfrac{\cos^2(\theta)}{b^2}}d\theta
	\end{align}
	where $a\geq1, b\geq 1$. If $a=b=1$ then $y^5$ vanishes and the surface represents a hyperboloid with spherical constant-time surfaces without any singularities.
	But if $b>1$ then the period of $\phi$ is less than $2\pi$, and conical singularities are present on the surface. The analysis of Einstein equations shows \cite{deserjackiw} that the parameter $b$ is related to the mass $M$ of a particle which resides in the singularity:
	\begin{align}
		b=\dfrac{1}{1-4GM},
	\end{align}
	where $G$ is a gravitational constant, and nonzero mass leads to the alteration of the period of $\phi$:
	\begin{align}
		y^a(\phi+2\pi(1-4GM))=y^a(\phi).
	\end{align}
	An interesting property of this solution is a necessary existence of a second particle in a so-called antipodal point. It has a clear geometrical explanation: it is indeed impossible to bend a sphere in such way that the only one singularity would appear on its surface, since in order to bend a sphere one needs to cut out a sector from it and match the edges of a cut to get a spindle-like surface \cite{fomenkosoros}. The parameters $a$ and $b$ play the role of bending parameters.
	\section{Explicit embeddings} \label{4}
	\subsection{Generalized G\"{o}del universe}
	Let us apply this method to the embedding of a so-called "Squashed AdS geometry" \cite{rooman} which can also be seen as generalized G\"{o}del universe. Its metric element has the form
	\begin{align}\label{godel}
		{ds^2}=d{t}^2+2\mu\sinh^2 {\chi} dt d\phi-d{\chi}^2- (\sinh^2 {\chi} + (1-\mu^2)\sinh^4 {\chi}) d\phi^2-dz^2
	\end{align}
	where G\"odel dimensional parameter is taken to be $R=1/2$. Here and hereafter $\mu>0$: $\mu^2=2$ corresponds to the Godel universe, whereas $\mu^2=1$ --- to a direct product of $\mathbb{R}$ and 2+1-dimensional AdS universe. The coordinate $z$ can be trivially embedded into 1-dimensional spacetime, so we will omit it and consider 2+1-dimensional factor space. 
	The symmetry group of the $2+1$-dimensional AdS spacetime is $SO(2,2)$. "Squashing" reduces it to $SO(2,1)\otimes SO(2)$. In addition to second $SO(2)$ associated with $\phi$ there is also an $SO(2)$ subgroup of $SO(2,1)$ associated with $t$.
	
	There are three metric components affected by these two Abelian subgroups: $g_{tt}$, $g_{\phi\phi}$ and  $g_{t\phi}$. Let us deal with the latter. Since $g_{t\phi}\neq f(t) g_{\phi\phi}$, \eqref{ura} will not work and we need to use \eqref{nond}. The component $g_{t\phi}$ therefore can be embedded via the following spacelike block:
	\begin{align}\begin{split}\label{godel1}
		&y^1 =  \sqrt{\frac{\mu}{a}}\sinh \chi \sin (\phi-a t),\\
		&y^2 =  \sqrt{\frac{\mu}{a}}\sinh \chi \cos (\phi-a t).\end{split}
	\end{align}
	Next metric component is $g_{\phi\phi}$, so we use \eqref{alg}. Its timelike embedding has the form 
	\begin{align}\begin{split}\label{godel2}
		&y^3=\sinh\chi\sqrt{\frac{\mu}{a}-1+(\mu^2-1)\sinh^2\chi}\sin (\phi),\\
		&y^4=\sinh\chi\sqrt{\frac{\mu}{a}-1+(\mu^2-1)\sinh^2\chi}\cos (\phi).\end{split}
	\end{align}
	where $a<\mu$.
	We see that none of the remaining metric components depend on $t$. We then need to embed $g_{tt}$. If we suppose that Abelian $t$-symmetry of the metric corresponds to a compact subgroup $SO(2)$, we should choose a timelike trigonometric form of its embedding \eqref{alg} (although a hyperbolic one is also possible, it does not lead to a global embedding, see \cite{2004.05882} for details):
	\begin{align}\begin{split}\label{godel3}
		&y^5=\dfrac{1}{b}\sqrt{1+a\mu\sinh^2\chi}\sin b t,\\
		&y^6=\dfrac{1}{b}\sqrt{1+a\mu\sinh^2\chi}\cos b t.\end{split}
	\end{align}
	The remaining metric component, $\tilde{g}_{\chi\chi}$, can be embedded into one-dimensional space using \eqref{int}:
	\begin{align}\label{godel4}
		y^7=f(\chi).
	\end{align}
	
	The explicit expression of $f(\chi)$ and the values of constants $a$ and $b$ can be found in \cite{2004.05882}. In the particular case when $a=\mu$, it gives an embedding described in \cite{rooman}.

	Now let us deal with another modification of 2+1-dimensional AdS spacetime, namely BTZ black hole.
	\subsection{BTZ}
	Let us consider the BTZ line element in the Eddington-Finkelstein coordinates\cite{Chan_1996}:
	\begin{align}\label{41}
		ds^2=\left(-M+\dfrac{r^2}{l^2}\right)dv^2-2dvdr+Jdvd\phi-r^2d\phi^2
	\end{align}
	In the static case ($J=0$) this metric can be globally embedded both in 5-dimensional \cite{Willison_2011} and 6-dimensional \cite{statja56} ambient spacetime, but the embedding of spinning BTZ black hole is much more tricky. Willison obtained \cite{Willison_2011} 10-dimensional embedding of the outer region, but, up to our knowledge, a global embedding of \eqref{41} is absent in the literature.\footnote{It should be noted that the embedding given in the original paper\cite{BTZ2} does not describe the metric \eqref{41} since in \cite{BTZ2} $\phi$ coordinate is non-compact and requires artificial compactification.} Let us construct it using the described method.    
	
	To get started, let us make a coordinate transformation:
	\begin{align}
		\phi=\theta+\frac{2}{J}r.
	\end{align}
	In these new coordinates, the metric takes the form
	\begin{align}
		ds^2=\left(-M+\dfrac{r^2}{l^2}\right)dv^2+Jdvd\theta-\dfrac{4r^2}{J^2}dr^2- \dfrac{4r^2}{J}drd\theta-r^2d\theta^2.
	\end{align}
	At the first glance, the transformation did not help us at all as it has not changed the number of non-diagonal metric components. They are, however, became easier to handle. Let us firstly embed $g_{v\theta}$ using \eqref{nond}: 
	\begin{align}\label{btz1}\begin{split}
			&y^1=\sqrt{\dfrac{J}{2a}}\sin\left(av+\theta\right),\\
			&y^2=\sqrt{\dfrac{J}{2a}}\cos\left(av+\theta\right),\\
		\end{split}
	\end{align}
	where signature is $(+,+)$. Note that $g_{\theta\theta}$ and $g_{vv}$ will be altered by this step.
	Then, since both $g_{\theta r}$ and $g_{\theta\theta}$ depends only on $r$ it is now possible to use \eqref{ura} and \eqref{h} in order to embed them at once using two spacelike components:
	\begin{align}\label{btz2}\begin{split}
			&y^3=\sqrt{r^2+\dfrac{J}{2a}}\sin\left(\theta+\dfrac{2}{J}r-\sqrt{\dfrac{2}{Ja}} \arctan \left(\sqrt{\dfrac{2a}{J}}r\right)\right),\\
			&y^4=\sqrt{r^2+\dfrac{J}{2a}}\cos\left(\theta+\dfrac{2}{J}r-\sqrt{\dfrac{2}{Ja}} \arctan \left(\sqrt{\dfrac{2a}{J}}r\right)\right).	\end{split}	
	\end{align}
	The remaining 2-dimensional surface with diagonal metric
	\begin{align}\label{btz_tilde}
		d\tilde{s}^2 = \left(-M+\dfrac{r^2}{l^2}-\dfrac{Ja}{2} \right)dv^2+\dfrac{r^2(2Ja-4)}{J(2ar^2+J)}dr^2
	\end{align}
	and coordinates $(v,r)$ has translational invariance w.r.t. $v$ and can therefore be embedded in a 3-dimensional ambient space by many ways described in Section \eqref{2}. Let us choose spiral embedding \eqref{spiral} with the signature $(-,+,+)$:
	\begin{align}\label{btz3}\begin{split}
			&y^5=b v+h(r),\\
			&y^6=f(r)\sin(c v+w(r)),\\
			&y^7=f(r)\cos(c v+w(r)).\end{split}
	\end{align}
	It is easy to check that $f(r)$ has the form
	\begin{align}
		f(r) = \frac{1}{c}\sqrt{b^2-M-\dfrac{Ja}{2}+\dfrac{r^2}{l^2}}.
	\end{align}
	Then these embedding function components \eqref{btz3} and the metric \eqref{btz_tilde} can be substituted in \eqref{metric}, which becomes a system of ODEs. From this system it can be seen that 
	the following reparametrization of $a$ turns out to be more convenient:
	\begin{align}
		a=\dfrac{2\alpha^2}{J},
	\end{align}
the functions $h(r)$ and $w(r)$ can be set to zero:
\begin{align}
		 w(r)=h(r)=0.
\end{align}	
This is not the most general form of them, but definitely the most simple one. In this case the remaining constants have the form
	\begin{align}
		b=\sqrt{\alpha^2+M+\dfrac{J^2}{4\alpha^2 l^2}},\quad c=\dfrac{1}{l}\sqrt{\dfrac{\alpha^2}{\alpha^2-1}}.
	\end{align}
Transforming the solutions back to the Eddington-Finkelstein coordinates, one obtains a remarkably simple embedding:
	\begin{align}\label{sol_btz}\begin{split}
			&y^1={\dfrac{J}{2\alpha}}\sin\left({\varphi}+\dfrac{2}{J}(\alpha^2 v-r)\right),\\
			&y^2={\dfrac{J}{2\alpha}}\cos\left({\varphi}+\dfrac{2}{J}(\alpha^2 v-r)\right),\\
			&y^3=\sqrt{r^2+\dfrac{J^2}{4\alpha^2}}\sin\left({\varphi}-\dfrac{1}{\alpha}{\arctan\left(\frac{2\alpha r}{J}\right)}\right),\\
			&y^4=\sqrt{r^2+\dfrac{J^2}{4\alpha^2}}\cos\left({\varphi}-\dfrac{1}{\alpha}{\arctan\left(\frac{2\alpha r}{J}\right)}\right),\\
			&y^5=v\sqrt{\alpha^2+M+\dfrac{J^2}{4\alpha^2 l^2}},\\
			&y^6=\dfrac{1}{\alpha}\sqrt{(\alpha^2-1)\left(r^2+\dfrac{J^2}{4\alpha^2}\right)}\sin\left(\dfrac{\alpha v}{l\sqrt{\alpha^2-1}}\right),\\
			&y^7=\dfrac{1}{\alpha}\sqrt{(\alpha^2-1)\left(r^2+\dfrac{J^2}{4\alpha^2 }\right)}\cos\left(\dfrac{\alpha v}{l\sqrt{\alpha^2-1}}\right),\end{split}
	\end{align}
	where $\alpha>1$ and signature is $(+ + - - - + +)$. As can be seen, it is smooth at all values of $v$, $\phi$ and $r$. It is especially interesting that this embedding remains smooth after the continuation to $r\leq 0$ (which is possible due to global properties of BTZ metric \cite{BTZ2}). 
	
	\section{Conclusion}\label{5}
	Let us put together the results of this paper. We propose an algorithm for construction of explicit embeddings of pseudo-Riemannian manifolds with a symmetry group. This algorithm helps to separate variables in the system of PDEs \eqref{metric} defining the embedding function, and consists of the following steps:
	\begin{enumerate}
		\item Identify the symmetry group of the metric and all of its Abelian subgroups.
		\item Find a coordinate system in which the metric components are independent of coordinates associated with all (or almost all) of these subgroups (i.e. looks like \eqref{genmet}).
		\item Embed non-diagonal components of the metric using \eqref{ura} or \eqref{nond} depending of their form.
		\item Embed diagonal components using \eqref{alg} and choosing the signature according to global properties of corresponding group.
		\item When the dependence of $y^a$ on all but one coordinate is determined, embed the remaining metric component using \eqref{int}. If the obtained embedding has some undesired properties, then return to the step when there is one symmetric coordinate left and embed the remaining 2-dimensional surface into ambient space using the original method (e.g. by \eqref{spiral}).
		Then substitute this embedding function and remaining metric into \eqref{metric} and solve the system of ODEs. 
		\item Determine the values of possible bending parameters in order to ensure that embedding function $y^a$ is real-valued.
	\end{enumerate}
	To illustrate the properties and advantages of this method, we have constructed explicit embeddings of 2+1 $\Lambda>0$ spacetime with point sources \eqref{L>0 emb}-\eqref{p}, G\"odel universe \eqref{godel1}-\eqref{godel4} and BTZ black hole \eqref{sol_btz}. The embedding of BTZ spacetime obtained here is global and relatively low-dimensional comparing to known ones. The construction of  embedding for spinning black hole in 2+1 dimensions might be an important step on the way to the construction of Kerr metric embedding.
	
	As a final remark, we want to emphasize that the method described here can be used in the construction of embeddings in non-flat spaces. The only alteration one should made is in the Poincare group of motion of ambient space, which needs to be replaced by a group of motion of the desired space. An embedding of Abelian subgroups through configurations which are more complicated than $SO(2)$ and $SO(1,1)$-rotations \eqref{alg} is also possible.  
	
	{\bf Acknowledgments.} {The authors are grateful to A. Starodubtsev for useful discussions. The work is supported by RFBR Grant No.~20-01-00081.}


\begin{thebibliography}{10}
	\newcommand{\enquote}[1]{``#1''}
	\providecommand{\url}[1]{\texttt{#1}}
	\providecommand{\urlprefix}{URL }
	\expandafter\ifx\csname urlstyle\endcsname\relax
	\providecommand{\doi}[1]{doi:\discretionary{}{}{}#1}\else
	\providecommand{\doi}{doi:\discretionary{}{}{}\begingroup
		\urlstyle{rm}\Url}\fi
	\providecommand{\eprint}[1]{\href{http://arxiv.org/abs/#1}{\texttt{#1}}}
	
	\bibitem{gauss1902}
	C.~F. Gauss, \href{https://archive.org/details/generalinvestiga00gausuoft/}{\enquote{General investigations of curved surfaces of
		1827 and 1825}}, University Library, [Princeton, 1902.
	
	\bibitem{Darrigol}
	O.~Darrigol, \enquote{The mystery of Riemann's curvature},
	\href{http://dx.doi.org/https://doi.org/10.1016/j.hm.2014.03.001}{\emph{Historia
			Mathematica}}, \textbf{42}: 1 (2015), 47--83.
	
	\bibitem{taimanov2008lectures}
	I. A. Taimanov, \enquote{Lectures on Differential Geometry}, EMS series of
	lectures in mathematics, European Mathematical Society, 2008.
	
	\bibitem{Phillips}
	E.~R. Phillips, \enquote{Karl M. Peterson: The earliest derivation of the
		Mainardi-Codazzi equations and the fundamental theorem of surface theory},
	\href{http://dx.doi.org/https://doi.org/10.1016/0315-0860(79)90075-2}{\emph{Historia
			Mathematica}}, \textbf{6}: 2 (1979), 137--163.
	
	\bibitem{Goenner1977}
	H.~F. Goenner, \enquote{On the interdependency of the Gauss-Codazzi-Ricci
		equations of local isometric embedding},
	\href{http://dx.doi.org/10.1007/BF00770733}{\emph{General Relativity and
			Gravitation}}, \textbf{8}: 2 (1977), 139--145.
	
	\bibitem{slemrod}
	M.~Slemrod, \enquote{Lectures on the Isometric Embedding Problem $(M^n, g) \to
		\mathbb{R}^m, m=\frac{n}{2}(n+1)$}, in \emph{Differential Geometry and
		Continuum Mechanics}, edited by Gui-Qiang~G. Chen, M. Grinfeld, R.~J.
	Knops, 77--120, Springer International Publishing, Cham, 2015.
	
	\bibitem{schmutzer}
	H.~Stephani, D.~Kramer, M.~Maccallum, C.~Hoenselaers, E.~Herlt, \enquote{Exact
		Solutions of Einstein's Field Equations, 2nd ed.}, Cambridge University
	Press, 2003.
	
	\bibitem{hanhong}
	Qing {Han}, Jia-Xing {Hong}, \enquote{{Isometric embedding of Riemannian
			manifolds in Euclidean spaces}}, \emph{{Math. Surv. Monogr.}}, vol. 130,
	Providence, RI: American Mathematical Society (AMS), 2006.
	
	\bibitem{Goenner1975}
	H. Goenner, \enquote{Local isometric embedding of riemannian manifolds with
		groups of motion}, \href{http://dx.doi.org/10.1007/BF00766605}{\emph{Gen.
			Rel. Grav.}}, \textbf{6}: 1 (1975), 75--78.
	
	\bibitem{Chantry_2021}
	L. Chantry, F. Dauvergne, Y. Temmam, V.
	Cayatte, \enquote{Quasi-isometric embedding of Kerr poloidal submanifolds},
	\href{http://dx.doi.org/10.1088/1361-6382/ac08a6}{\emph{Classical and Quantum
			Gravity}}, \textbf{38}: 14 (2021), 145030.
	
	\bibitem{statja27}
	S.~A. Paston, A.~A. Sheykin, \enquote{Embeddings for Schwarzschild metric:
		classification and new results},
	\href{http://dx.doi.org/10.1088/0264-9381/29/9/095022}{\emph{Class. Quant.
			Grav.}}, \textbf{29} (2012), 095022, \eprint{arXiv:1202.1204}.
	
	\bibitem{statja30}
	S.~A. Paston, A.~A. Sheykin, \enquote{Global embedding of the
		Reissner-Nordstrom metric in the flat ambient space},
	\href{http://dx.doi.org/10.3842/SIGMA.2014.003}{\emph{SIGMA}}, \textbf{10}
	(2014), 003, \eprint{arXiv:1304.6550}.
	
	\bibitem{statja32}
	A.~A. Sheykin, D.~A. Grad, S.~A. Paston, \enquote{Embeddings of the black holes
		in a flat space}, in \emph{Proceedings of QFTHEP 2013, Saint Petersburg Area,
		Russia}, \href{https://doi.org/10.22323/1.183.0091}{Proceedings of Science}, PoS(QFTHEP2013)091, \eprint{arXiv:1401.7820}.
	
	\bibitem{statja40}
	A.~A. Sheykin, S.~A. Paston, \enquote{Classification of minimum global
		embeddings for nonrotating black holes},
	\href{http://dx.doi.org/10.1007/s11232-015-0364-1}{\emph{Theor. Math.
			Phys.}}, \textbf{185}: 1 (2015), 1547--1556, \eprint{arXiv:1512.08280}.
	
	\bibitem{statja57}
	A.~D. Kapustin, M.~V. Ioffe, S.~A. Paston, \enquote{Explicit isometric
		embeddings of collapsing dust ball},
	\href{http://dx.doi.org/10.1088/1361-6382/ab74f8}{\emph{Classical and Quantum
			Gravity}}, \textbf{37}: 7 (2020), 075019, \eprint{arXiv:2003.03742}.
	
	\bibitem{statja56}
	A.~A. Sheykin, D.~P. Solovyev, S.~A. Paston, \enquote{Global Embeddings of BTZ
		and Schwarzschild-ADS Type Black Holes in a Flat Space},
	\href{http://dx.doi.org/10.3390/sym12020240}{\emph{Symmetry}}, \textbf{11}: 7
	(2019), 841, \eprint{arXiv:1905.10869}.
	
	\bibitem{statja29}
	S.~A. Paston, A.~A. Sheykin, \enquote{Embeddings for solutions of Einstein
		equations}, \href{http://dx.doi.org/10.1007/s11232-013-0067-4}{\emph{Theor.
			Math. Phys}}, \textbf{175}: 3 (2013), 806--815, \eprint{arXiv:1306.4826}.
	
	\bibitem{2004.05882}
	A.~A. Sheykin, M.~V. Markov, Ya.~A. Fedulov, S.~A. Paston, \enquote{Explicit
		isometric embeddings of pseudo-Riemannian manifolds: ideas and applications},
	\href{http://dx.doi.org/10.1088/1742-6596/1697/1/012077}{\emph{Journal of
			Physics: Conference Series}}, \textbf{1697} (2020), 012077,
	\eprint{2004.05882}.
	
	\bibitem{Willison_2011}
	S. Willison, \enquote{The Banados, Teitelboim, and Zanelli spacetime as an
		algebraic embedding},
	\href{http://dx.doi.org/10.1063/1.3579486}{\emph{J. Math. Phys.}}, \textbf{52}: 4 (2011), 042503, \eprint{1011.3883}.
	
	\bibitem{gane}
	M.~Janet, \enquote{Sur la possibilite de plonger un espace riemannien donne
		dans un espace euclidien}, \emph{Ann. Soc. Polon. Math.}, \textbf{5} (1926),
	38--43.
	
	\bibitem{burstin}
	C.~{Burstin}, \enquote{{Beitr\"age der Verbiegung von Hyperfl\"achen in
			euklidischen R\"aumen}}, \emph{{Rec. Math. Moscou}}, \textbf{38}: 3-4 (1931),
	86--93.
	
	\bibitem{kart}
	E.~Kartan, \enquote{Sur la possibilite de plonger un espace riemannien donne
		dans un espace euclidien}, \emph{Ann. Soc. Polon. Math.}, \textbf{6} (1927),
	1--7.
	
	\bibitem{fridman61}
	A.~Friedman, \enquote{Local isometric embedding of Riemannian manifolds with
		indefinite metric},
	\href{http://dx.doi.org/10.1512/iumj.1961.10.10042}{\emph{J. Math. Mech.}},
	\textbf{10} (1961), 625.
	
	\bibitem{Gromov_1970}
	M.~L. Gromov, V.~A. Rokhlin, \enquote{Embeddings and immersions in Riemannian
		geometry},
	\href{http://dx.doi.org/10.1070/rm1970v025n05abeh003801}{\emph{Russian
			Math. Surveys}}, \textbf{25}: 5 (1970), 1--57.
	
	\bibitem{statja25}
	S.~A. Paston, \enquote{Gravity as a field theory in flat space-time},
	\href{http://dx.doi.org/10.1007/s11232-011-0138-3}{\emph{Theor. Math.
			Phys.}}, \textbf{169}: 2 (2011), 1611--1619, \eprint{arXiv:1111.1104}.
	
	\bibitem{statja61}
	S.~A. Paston, E.~N. Semenova, A.~A. Sheykin, \enquote{Canonical description for
		formulation of embedding gravity in a form of field theory in a flat
		spacetime}, \href{http://dx.doi.org/10.3390/sym12050722}{\emph{Symmetry}},
	\textbf{12}: 5 (2020), 722, \eprint{arXiv:2004.04481}.
	
	\bibitem{jensen}
	P.~A. Griffiths, G.~R. Jensen,\href{http://www.jstor.org/stable/j.ctt1b9x2fk}{ \enquote{Differential Systems and
		Isometric Embeddings}}, Princeton University Press, 1987.
	
	\bibitem{Hall2015}
	B.~C. Hall, \href{https://doi.org/10.1007/978-3-319-13467-3_1}{Matrix Lie Groups}, 3--30, Springer International Publishing,
	Cham, 2015.
	
	\bibitem{vilenkin}
	N.~Ya. {Vilenkin}, \href{http://www.mathnet.ru/php/archive.phtml?wshow=paper&jrnid=sm&paperid=4325&option_lang=rus}{\enquote{{Polyspheric and orispheric functions (\textit{in
				Russian})}}}, \emph{{Mat. Sb., Nov. Ser.}}, \textbf{68} (1965), 432--443.
	
	\bibitem{letaw81}
	J.~R. Letaw, \enquote{Stationary world lines and the vacuum excitation of
		noninertial detectors},
	\href{http://dx.doi.org/10.1103/PhysRevD.23.1709}{\emph{Phys. Rev. D}},
	\textbf{23} (1981), 1709--1714.
	
	\bibitem{deserjackiw}
	S.~Deser, R.~Jackiw, \enquote{Three-dimensional cosmological gravity: Dynamics of
		constant curvature},
	\href{http://dx.doi.org/https://doi.org/10.1016/0003-4916(84)90025-3}{\emph{Annals
			of Physics}}, \textbf{153}: 2 (1984), 405--416.
	
	\bibitem{fomenkosoros}
	V.~T. Fomenko, \href{http://web.archive.org/web/20030703003304/http://www.issep.rssi.ru/pdf/9805_122.pdf}{\enquote{Bending of surfaces (in Russian)}}, \emph{Soros. Obr.
		J.}, \textbf{5} (1998), 122--127.
	
	\bibitem{rooman}
	M.~Rooman, Ph.~Spindel, \enquote{Godel metric as a squashed anti-de Sitter
		geometry},
	\href{http://dx.doi.org/10.1088/0264-9381/15/10/024}{\emph{Classical and
			Quantum Gravity}}, \textbf{15}: 10 (1998), 3241--3249,
	\eprint{gr-qc/9804027}.
	
	\bibitem{Chan_1996}
	J.~S.~F. Chan, K.~C.~K. Chan, R.~B. Mann, \enquote{Interior structure of a
		spinning black hole in 2+1 dimensions},
	\href{http://dx.doi.org/10.1103/physrevd.54.1535}{\emph{Physical Review D}},
	\textbf{54}: 2 (1996), 1535--1539, \eprint{gr-qc/9406049}.
	
	\bibitem{BTZ2}
	M. Ba\~nados, M. Henneaux, C. Teitelboim, J. Zanelli,
	\enquote{Geometry of the 2+1 black hole},
	\href{http://dx.doi.org/10.1103/PhysRevD.48.1506}{\emph{Phys. Rev. D}},
	\textbf{48}: 4 (1993), 1506--1525.
	
\end{thebibliography}
\end{document}